\documentclass[prb,twocolumn,showpacs,aps]{revtex4}
\usepackage{color,soul}
\usepackage {graphicx}
\usepackage {epstopdf}

\begin{document}
\title{Evolution of magnetic and transport properties in Cu doped pyrochlore iridate Eu$_{2}$(Ir$_{1-x}$Cu$_{x}$)$_{2}$O$_{7}$}

\author{Sampad Mondal,$^{a,b,c}$\footnote{Email:sampad100@gmail.com} M. Modak,$^b$ B. Maji,$^d$ Swapan K. Mandal,$^a$ B. Ghosh,$^e$ Surajit Saha,$^e$ M. Sardar$^f$ and S. Banerjee$^b$\footnote{Email:sangam.banerjee@saha.ac.in}}
\address{$^a$Department of Physics, Visva-Bharati, Santiniketan 731235, India\\
	$^b$Saha Institute of Nuclear Physics, 1/AF Bidhannagar, Kolkata 700064, India\\
	$^c$Ramsaday College, Amta, Howrah 711401, India\\
	$^d$Acharya Jagadish Chandra Bose College, 1/1B, A. J. C. Bose Road, Kolkata 700020, India\\
	$^e$Department of Physics, Indian Institute of Science Education and Research, Bhopal 462066, India\\
	$^f$Material Science Division, Indira Gandhi Centre for Atomic Research, Kalpakkam 603102, India}
\begin{abstract}
  We have investigated the effect of Cu substitution in Eu$_{2}$(Ir$_{1-x}$Cu$_{x}$)$_{2}$O$_{7}$ with the help of magnetic and transport property measurements.
  XPS measurement reveals that each Cu$^{2+}$ converts Ir$^{4+}$ to double amount of Ir$^{5+}$ ions. The metal-insulator transition temperature (T$_{MI}$) is obtained around 120 K. In the insulating phase, at lower temperature  below 50 K, the temperature dependent resistivity follows a power law dependence and the magnitude of the exponent increases with Cu concentrations. The temperature dependent thermopower is observed to follow the electrical resistivity down to 50 K, except for a sudden drop in thermopower at temperature below 50 K. We find negligible Hall voltage in the metallic regime of the samples but a sudden Hall voltage is developed below 50 K. We observe bifurcation in zero field cooled and field cooled (ZFC-FC) magnetization below irreversibility temperature, exchange bias and negative magnetoresistance at 3 K and the magnitude of all these properties increases with Cu concentrations. In the insulating region (below 6 K) there exists a linear specific heat and its coefficient decreases with Cu doping which indicates the reduction of spinon contribution with Cu doping.
  
\end{abstract}
\maketitle
{\large\bf{Introduction}}
\vskip 0.2cm
 Unlike $3d$ and $4d$ transition metal oxides, there exists a relativistic spin-orbit interaction in Hamiltonian in $5d$ transition metal oxide compounds. The competition of spin orbit interaction and electronic correlation energy U, gives rise to a rich variety of novel quantum correlated phases such as quantum spin liquid, Weyl semi-metal and Axion insulator.\cite{MacLaughlin,Wan,Yamaji,Witczek-Krempa} In $5d$ transition metal oxide compound, for example, layered Iridium oxide Sr$_2$IrO$_4$ system\cite{Kim1}, Ir is in charge state Ir$^{4+}$ ($5d^5$). The crystal field effect due to the environment of IrO$_6$ octahedra, splits the $5d$ orbital into t$_{2g}$ and e$_{g}$ levels. In presence of spin orbit coupling (SOC) the lower energy level t$_{2g}$ further splits into two bands with fully filled spin J$_{eff}$=3/2 quadruplet and partially filled spin J$_{eff}$=1/2 doublet. The electron correlation further splits the J$_{eff}$=1/2 band and opens a Mott gap at the Fermi level (E$_F$). Thus the interplay of SOC and U provides a stage for the study of a wide range of quantum correlated phenomena.  

Among the $5d$ transition metal oxide compounds, pyrochlores are the most promising system to tune SOC and U because of their interpenetrating corner sharing tetrahedral structure. It tends to form a narrow flat band at Fermi level and the band width is comparable to energy scale of SOC and U. Particularly, R$_2$Ir$_2$O$_7$ (R=rare earth) shows an intriguing behavior in their exotic transport and magnetic properties depending on the ionic radius of the rare earth atom.\cite{Tian,MATSUHIRA,Daiki} In the pyrochlore family, R=Nd-Sm shows metal-insulator transition with an all-in/all-out antiferromagnetic spin ordering at insulating state. Recently, it has been theoretically predicted that this compound shows novel correlated quantum phenomena including topological insulator, Weyl semimetal, Axion insulator.\cite{pesin,yang,Wan,Zhang} In  a magnetically ordered system, where time reversal symmetry is broken but inversion symmetry is preserved, quadratic band touching point leads to a linear dispersing Dirac fermion like spectrum, along with definite spin chirality i.e., $\bar{k}.\bar{\sigma} =\pm$ 1, where $\bar{k}$ and $\bar{\sigma}$ are the unit vectors along the momentum and spin of the electron respectively. This promotes the formation of Weyl semimetal like phase in this compound. In the strong correlation limit, the pre-existing narrow bands due to spin orbit coupling may open up a gap. For an increase in Coulomb correlation energy, pair of Weyl point with opposite chirality moves towards Brillouin zone boundary and annihilates pairwise to open up a gap forming a spin orbit assisted Mott insulator.\cite{pesin} If there exists frustration in the magnetic exchange interaction (like in the pyrochlore compounds), then one may get either metallic or insulating spin liquid phase.\cite{nakatsuji}

To study such exotic ground state properties in details the pyrochlore iridate Eu$_{2}$Ir$_{2}$O$_{7}$ have drawn special interest. In such a compound, one can not only observe thermally induced metal-insulator transition but also can avoid $f-d$ exchange interaction \cite{Chen} for non-magnetic Eu$^{3+}$ ion. This allows us to study the Ir sublattice without the phenomena emerging from interaction with rare earth ion and Ir$^{4+}$ ion. Weyl semimetal signature for Eu$_{2}$Ir$_{2}$O$_{7}$ was confirmed by Tafti et al.\cite{Tafti} comparing with the calculated resistivity data\cite{Hosur} for Weyl semimetal. Recently, Telang et al.\cite{Telang} also reported the Weyl semimetal nature in the polycrystalline Eu$_{2}$Ir$_{2}$O$_{7}$ compound by doping Bi at Eu site. All the above features make this system ideal for studying the interplay between SOC and U associated to the system. Substituting Ir by $3d$ or $4d$ element, one can tune the SOC and electronic correlation energy which can modify the physical property of the system. \cite{Noor,Kumar} Previously, Banerjee et al.\cite{Banerjee} showed that non-magnetic divalent element doping at Eu site creates a Ir$^{4+}$/Ir$^{5+}$ charge disproportion, which led the system toward non-Fermi liquid state.  

 In our present work, we have studied the effect of Cu doping at Ir site in Eu$_{2}$(Ir$_{1-x}$Cu$_{x}$)$_{2}$O$_{7}$ compound. To be mentioned,  Ir$^{4+}$ have higher SOC and lower U compared to Cu$^{2+}$. Substitution of Cu$^{2+}$ in Ir$^{4+}$ site not only changes the SOC and U in the system but also changes the state of Ir$^{4+}$ to Ir$^{5+}$, acts as site dilution for creating non-magnetic Ir$^{5+}$ charge state. We observe insignificant change in T$_{MI}$ with Cu doping and T$_{MI}$=  T$_{irr}$ (irreversibility temperature) is maintained for all the compounds. Upon doping of Cu at Ir site, each Cu creates double amount of Ir$^{5+}$, thus each corner shared tetrahedron deviates from its all-in-all-out (AIAO) spin structure and picks up a moment, which leads to a small  ferro contribution in those droplet regions (Note: this supermoment droplet we shall refer to it as ferro contribution in this manuscript) dispersed within a background of antiferromagnetic all-in-all-out (AFM AIAO) spin structure. It is believed that quantum fluctuations could stabilize a so-called resonating valence bond (RVB) state. The resulting RVB state is a liquid-like state of spins.\cite{Anderson} The prerequisite condition for spin liquid formation is AFM spin structure and hence, with increase in Cu doping the spinon contribution (fractional excitation for the spin liquid) region is found to be decreased. In this report, we shall address this in more details and also, its manifestations in physical properties such as in magnetic, electrical and thermal behavior will be discussed. 
\vskip 0.5cm
{\large\bf{Experimental details}} 
\vskip 0.2cm
All the polycrystalline samples were prepared by solid state reaction method. High purity ingredient powder Eu$_{2}$O$_{3}$, IrO$_{2}$ and CuO with phase purity 99.9\% (Alfa Aesar) were mixed in stoichiometric ratio and grounded well. After pressing the mixture powder in pellet form, the sample was heated at 1273 K for 3 days with several intermediate grinding. All the samples were characterized by powder X-ray diffraction (XRD), X-ray photoelectron spectroscopy (XPS). The room temperature XRD measurement was carrried out by X-ray diffractometer with Cu K$_{\alpha}$ radiation. Structural parameters were determined using standard Rietveld technique with Fullprof software package. XPS measurements at room temperature were carried out by using an Omicron Multiprobe Electron Microscopy System equipped with a monochromatic Al K$_{\alpha}$ X-ray source ($h\nu$ = 1486.7 eV). All the XPS spectra have been analyzed by using PeakFit software, where Shirley method was used for background subtraction. Magnetic measurements were performed using Superconducting Quantum Interference Device Magnetometer (SQUID-VSM) of Quantum Design in the temperature range 3-300 K. Electrical, magnetic and thermal transport measurements were carried out by four probe method with temperature range 2-300 K using Physical Properties Measurement System (PPMS). To determine the Hall coefficient, we have measured the Hall resistance with applied fields of opposite polarities, and subtracted one from the other. 
\vskip 0.5cm
{\large\bf{Experimantal results}} 
\vskip 0.2cm
\begin{figure} 
	\centering
	\includegraphics[width= 8 cm]{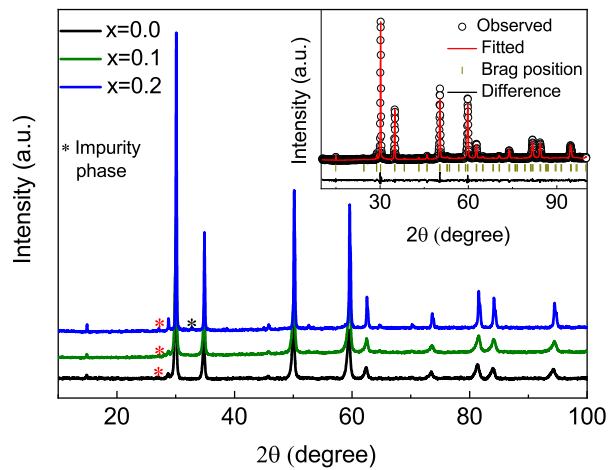}
	\caption{Room temperature X-ray diffraction pattern of  Eu$_{2}$(Ir$_{1-x}$Cu$_{x}$)$_{2}$O$_{7}$, where $x$=0, 0.1, 0.2 and red and black stars are indicating the impurity phases due to nonreacting IrO$_2$ and Eu$_2$O$_3$ respectively. Inset shows Rietveld refinement for $x$=0.1 compound, where scattered data are observed and solid line is fit to the data.}
	\label{XRD}
\end{figure}

Room temperature XRD pattern of Eu$_{2}$(Ir$_{1-x}$Cu$_{x}$)$_{2}$O$_{7}$ with $x$=0, 0.1, 0.2 are shown in fig. \ref{XRD}. XRD patterns of doped and parent samples appear similar. Rietveld refinement for all the compounds have been performed considering cubic structure with $Fd-3m$ space group, one of which ($x$=0.1) has been shown in the inset of fig. \ref{XRD}. All the samples are nearly in pure phase with some minor impurity phases (non-reacting oxide), indicated by red and black stars. The obtained lattice parameter for $x$=0, 0.1 and 0.2 compounds are 10.3048$\pm$0.0004, 10.2778$\pm$0.0001 and 10.2777$\pm$0.0001 \AA,  respectively. The lattice parameter decreases upon Cu doping. 

 \begin{figure*} 
 \centering
 \includegraphics[width= 17.5cm]{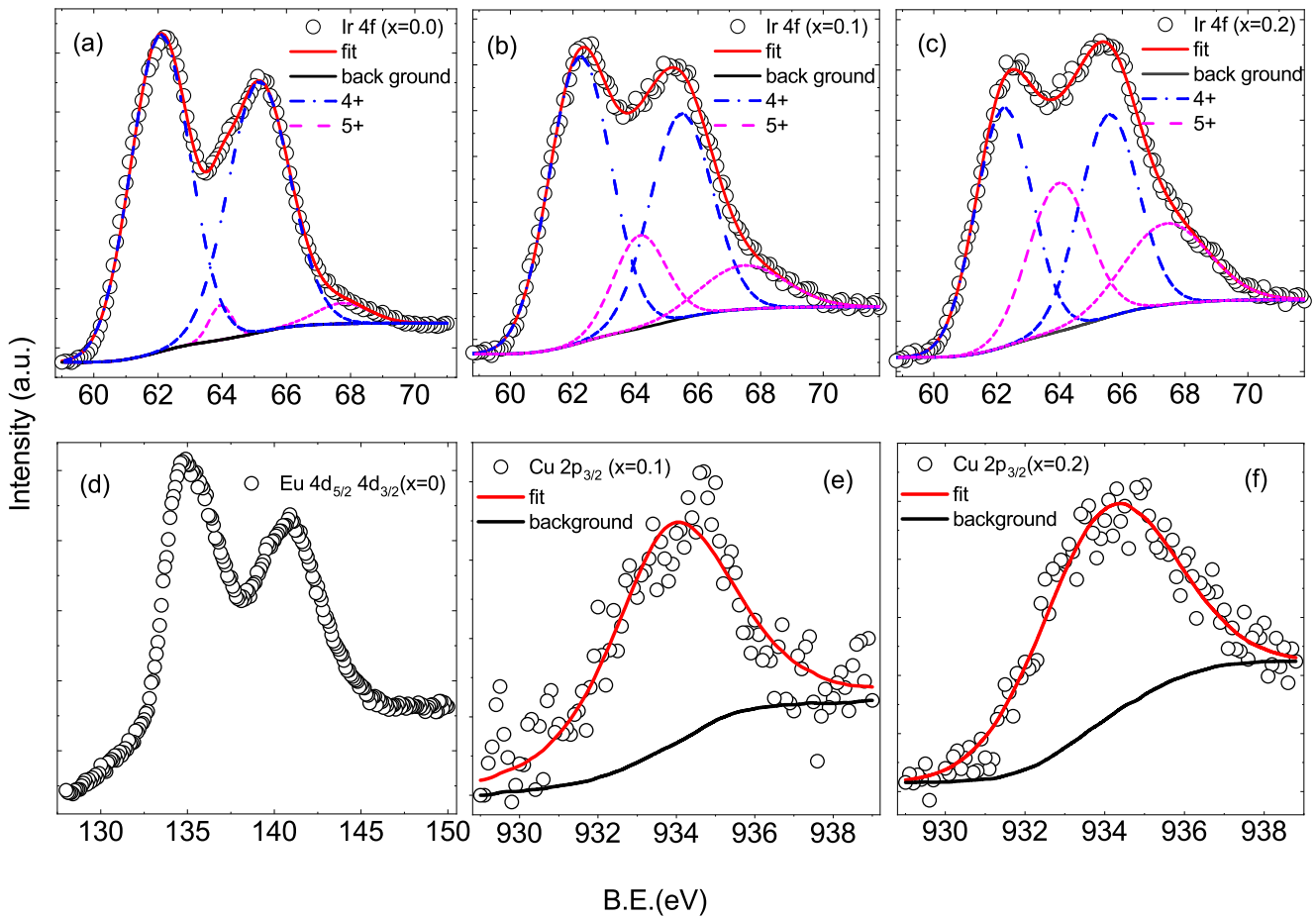}
 \caption{(a-c) Ir-4f core level X-ray photoelectrons spectra for $x$=0, 0.1, 0.2 compounds respectively. (d) Eu-4d spectra for $x$=0 and (e-f) Cu 2p$_{3/2}$ core level spectra for $x$=0.1, 0.2. Open black circle are experimental data and solid red lines are fitted data.}
 \label{XPS}
\end{figure*}
 
It is important to know the charge state of transition metal in this compound as it dominates the electronic and magnetic properties. Therefore, we have carried out X-ray photoelectrons spectroscopy (XPS) measurement for all the samples. Fig. \ref{XPS}(a-c) depicts the Ir 4f core level spectra for $x$=0, 0.1 and 0.2 respectively. Spin orbit coupling (SOC) splits the Ir 4f core level into  4f$_{7/2}$ and 4f$_{5/2}$ electronic states around binding energy 62 and 65 eV respectively (as shown by dashed blue line) for Ir$^{4+}$. The spectra for all the samples can be fitted taking the peaks due to Ir$^{4+}$ with some weak additional peaks occurring at the higher binding energy 64 and 67.7 eV (as shown by pink dotted line). These additional peaks correspond to 4f$_{7/2}$ and 4f$_{5/2}$ states of Ir$^{5+}$ ion. We observe a small increase in intensity of the Ir$^{5+}$ peak compared to Ir$^{4+}$ for $x$=0 compound. The calculation of area under the curve of the XPS spectrum reveals that for $x$=0 compound, the amount of Ir$^{4+}$  and Ir$^{5+}$ are 95.9\% and 4.1\% respectively. In the parent sample majority of Ir is in Ir$^{4+}$ state. The small amount of Ir$^{5+}$ in the Eu$_{2}$Ir$_{2}$O$_{7}$ could be due to unavoidable non-stoichiometry of the sample.\cite{Ishikawa,Zhu} We observe a significant increment in the intensity of Ir$^{5+}$ peaks compared to Ir$^{4+}$ in doped compounds and the amount of Ir$^{4+}$ and Ir$^{5+}$ in $x$=0.1, 0.2 compounds are found as 77.7\% and 22.3\%, 60.1\% and 39.9\% respectively.  Now if we consider Ir$^{5+}$ contribution due to the non-stoichiometry of the sample then 10\% and 20\% substitution of Cu produce roughly 18.2\% and 35.8\% Ir$^{5+}$ respectively. Hence, each Cu ion converts two Ir$^{4+}$ to Ir$^{5+}$ ions. Fig. \ref{XPS}(d) shows that Eu 4d$_{5/2}$ and 4d$_{3/2}$ core level spectra with peak position at 134.7 and 140.9 eV respectively. To further investigate the charge state of Cu in this compound, we have taken Cu 2p$_{3/2}$ core level spectra for $x$=0.1 and 0.2 respectively, shown in fig. \ref{XPS}(e-f). Red lines are fit to the observed data with 2p$_{3/2}$ spectra centered around the binding energy 934 eV, which confirm that Cu stays at Cu$^{2+}$ oxidization state. 

\begin{figure} 
	\centering
	\includegraphics[width= 8 cm]{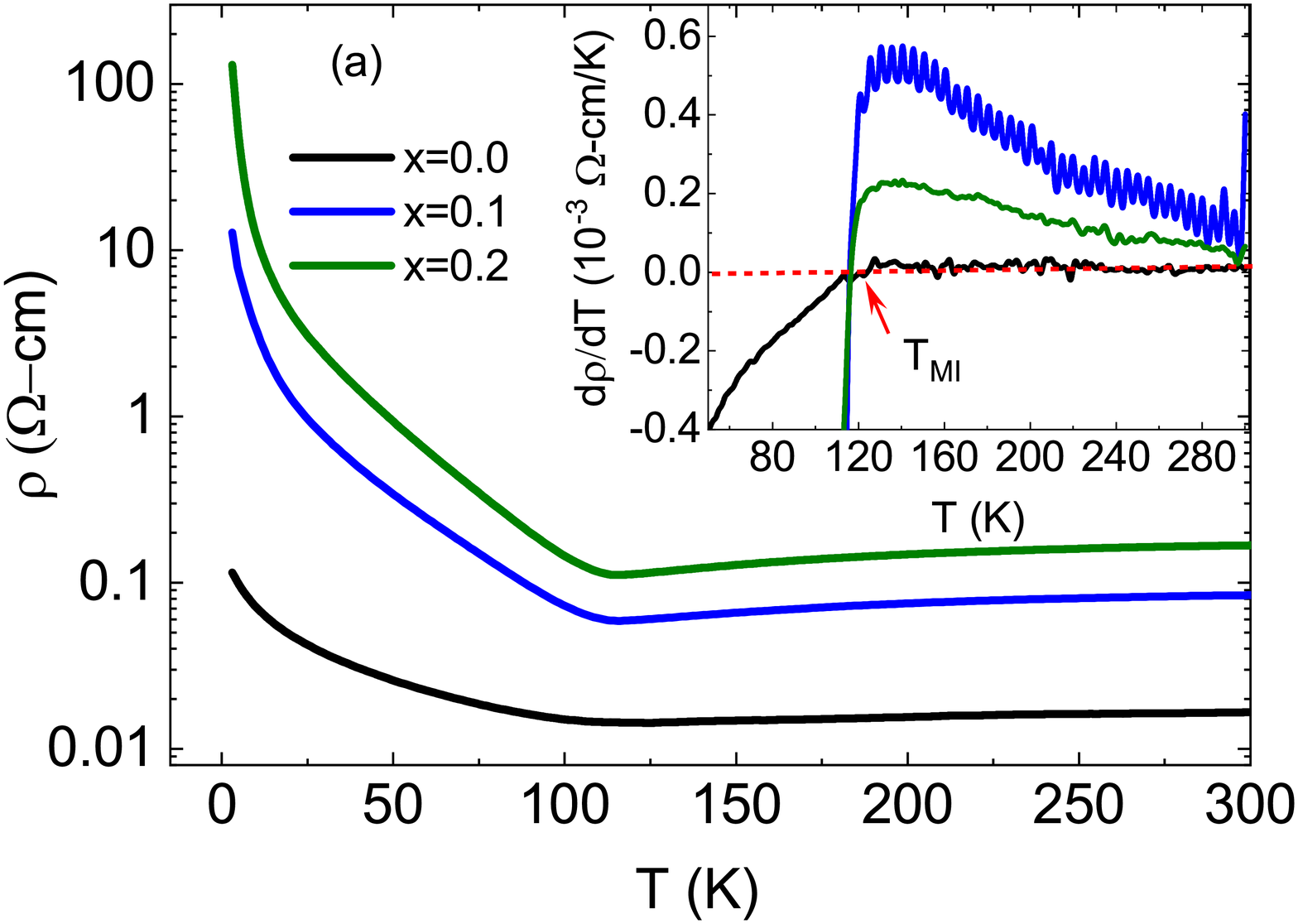}
	\includegraphics[width= 8cm]{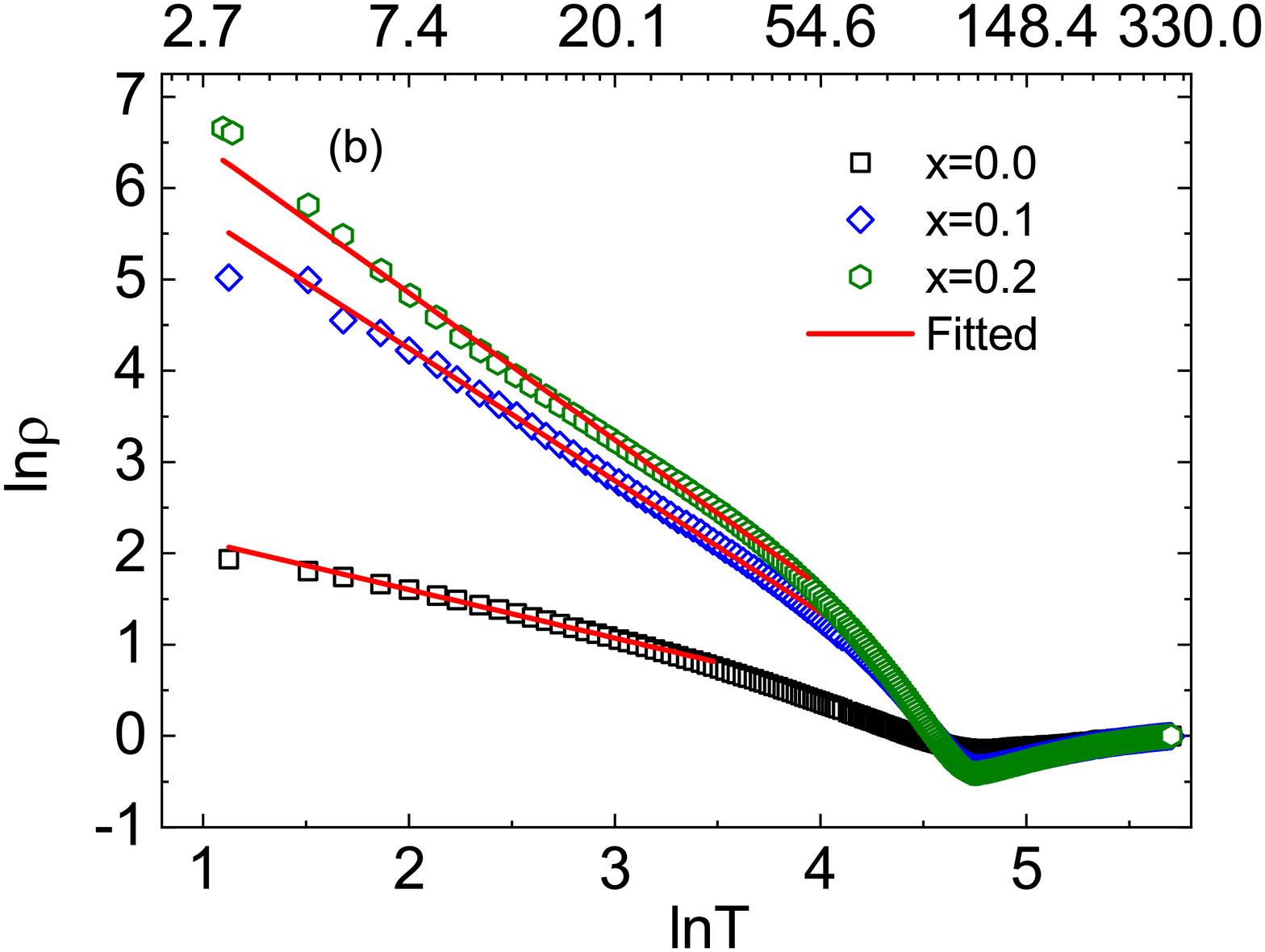}
	\caption{ (a) Temperature dependent resistivity measured in the temperature range 2.1 K- 300 K for Eu$_{2}$(Ir$_{1-x}$Cu$_{x}$)$_{2}$O$_{7}$. Inset: Temperature derivative of the resistivity as a function of temperature. (b) ln-ln plot of the $\rho $(T) data for all the compounds and solid red lines are power law fitting of the observed data.}
	\label{Rho-T}
\end{figure}
Resistivity measurements, carried out as a function of temperature, are presented in fig. \ref{Rho-T}(a). All the samples show metal to insulator transition (MIT), which is clearly observed from inset in fig. \ref{Rho-T}(a) where $d\rho/dT$ changes from positive to negative at lower temperature. Metal insulator transition temperatures (T$_{MI}$) for the compounds $x$=0, 0.1 and 0.2 are 120 K, 113 K and 115 K respectively. Cu doping shifts the T$_{MI}$ to lower temperature compared to the parent compound and the resistivity is higher than the parent compound in the entire temperature range.
\begin{table*}
	
	\centering \caption{Fitting parameters $\alpha$ obtained from Power law fitting for the samples Eu$_{2}$(Ir$_{1-x}$Cu$_{x}$)$_{2}$O$_{7}$.}

	\begin{tabular}{ccc}
		\hline
		\hline Sample &\hspace{0.5cm}Temperature range (K)&\hspace{0.5cm}$\alpha$\hspace{0.5cm}\\
		\hline $x$=0 & 4.5-23 &\hspace{0.5cm} 0.531$\pm$0.006\\
		\hline $x$=0.1 & 3-51 &\hspace{0.5cm} 1.446$\pm$0.007\\
		\hline $x$=0.2 & 5-49 & \hspace{0.5cm}1.61$\pm$0.01\\

		\hline
		\hline
	\end{tabular}
\end{table*}

For further analysis of the resistivity data at low temperature, we have tried to fit the data below T$_{MI}$ for all the samples with power law 
\begin{equation}
	\rho = \rho_{0}T^{-\alpha}
\end{equation}
where $\alpha$ is the exponent, as shown in fig. \ref{Rho-T}(b). The fitted parameter and temperature region are shown in table-I. All the data fits well over a limited temperature range (around one order of magnitude i.e., 3-50 K) and the $\alpha$ value is found to increase with Cu concentration. Power law behavior signifies that it is not an activated process and could be a continuous transition to a fluctuating disorder state. 

  In a 3-dimensional disordered system, Mott variable range hopping can be expected. Therefore, we have also tried to fit the resistivity data with three-dimensional Mott variable range hopping model (VRH)\cite{Mott}
\begin{equation}
	\rho = \rho_{0}\hspace{0.2cm}exp((T_{0}/T)^{1/4})
\end{equation}
where,  T$_{0}$ = $\frac{21.2}{N(E_{F}) \xi^{3}}$, $N(E_F)$  and $\xi$ are density of states at Fermi level and
localization length. All the samples can be fitted with a limited temperature region 3-10 K. So, we do not claim the system to be a strongly localized since the fitting temperature range is narrow (less than one order of magnitude).

\begin{figure} 
	\centering
	\includegraphics[width= 8 cm]{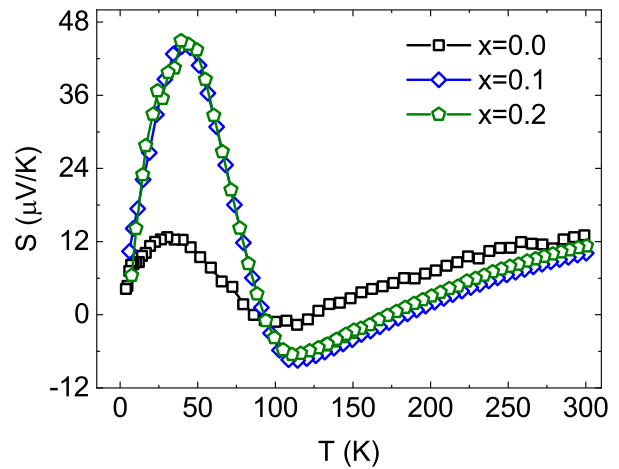}
	\caption{Temperature variation of thermoelectric power (S) for  Eu$_{2}$(Ir$_{1-x}$Cu$_{x}$)$_{2}$O$_{7}$, where $x$=0, 0.1, 0.2.}
	\label{S-T}
\end{figure}

  We have measured the temperature dependent Seebeck coefficient (S) as shown in fig. \ref{S-T}. The overall behavior of S(T) for $x$=0 is in agreement with the previous report.\cite{MATSUHIRA,Telang}  The nature of the temperature dependent S for doped samples is similar to the parent compound.  For all the samples, S is found to decrease monotonically from room temperature down to  T$_{MI}$, which is the expected behavior of a metal. On further reduction of temperature, S starts rising and at lower temperature below 50 to 30 K, a sudden drop is observed giving rise to a peak. This characteristic peak is possibly due to phonon-drag contribution as discussed in the previous reports.\cite{Telang1,MacDonald,MATSUHIRA,Telang} However, in this temperature range, the resistivity increases obeying power law behavior with reduction in temperature. This plausibly suggests a spin liquid like signature that will be explored later from the low temperature heat capacity measurement.
   
\begin{figure} 
	\centering
	\includegraphics[width= 8 cm]{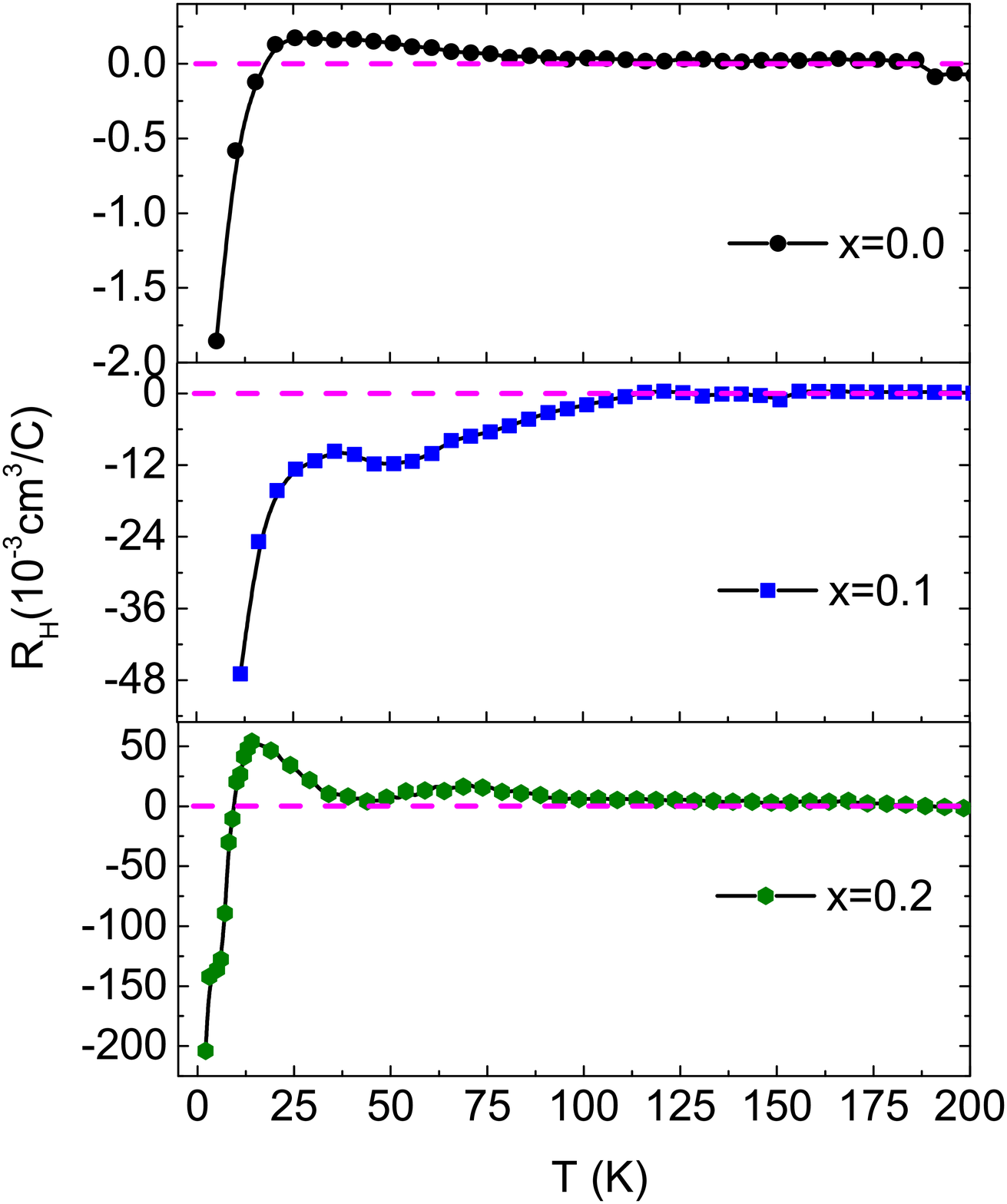}
	\caption{Temperature dependent Hall coefficient with applied field 50 kOe for  Eu$_{2}$(Ir$_{1-x}$Cu$_{x}$)$_{2}$O$_{7}$, where $x$=0, 0.1 and 0.2.}
	\label{hall}
\end{figure}

 We have also measured the Hall coefficient (proportional to the measured Hall voltage) as a function of temperature, with applied field 50 kOe for the parent and doped compounds, as shown in fig. \ref{hall}. 
 Hall voltage is negligible above
 T$_{MI}$ but below 50 to 30 K, a steep negative Hall voltage is found to be developed. We have also observed a drop in thermopower in the very similar temperature range (see fig. \ref{S-T}). The sudden development of the negative Hall voltage is possibly due to the onset of electron spin fluctuation affected by the applied perpendicular magnetic field.  Although, all the compounds show negative Hall coefficient (R$_H$) at low temperature, its magnitude for doped sample is observed to be higher than the parent sample. It indicates that the majority charge carriers at low temperature for all the compounds are electron-like, whereas the carrier concentration decreases with Cu doping.
\begin{figure} 
	\centering
	\includegraphics[width= 8 cm]{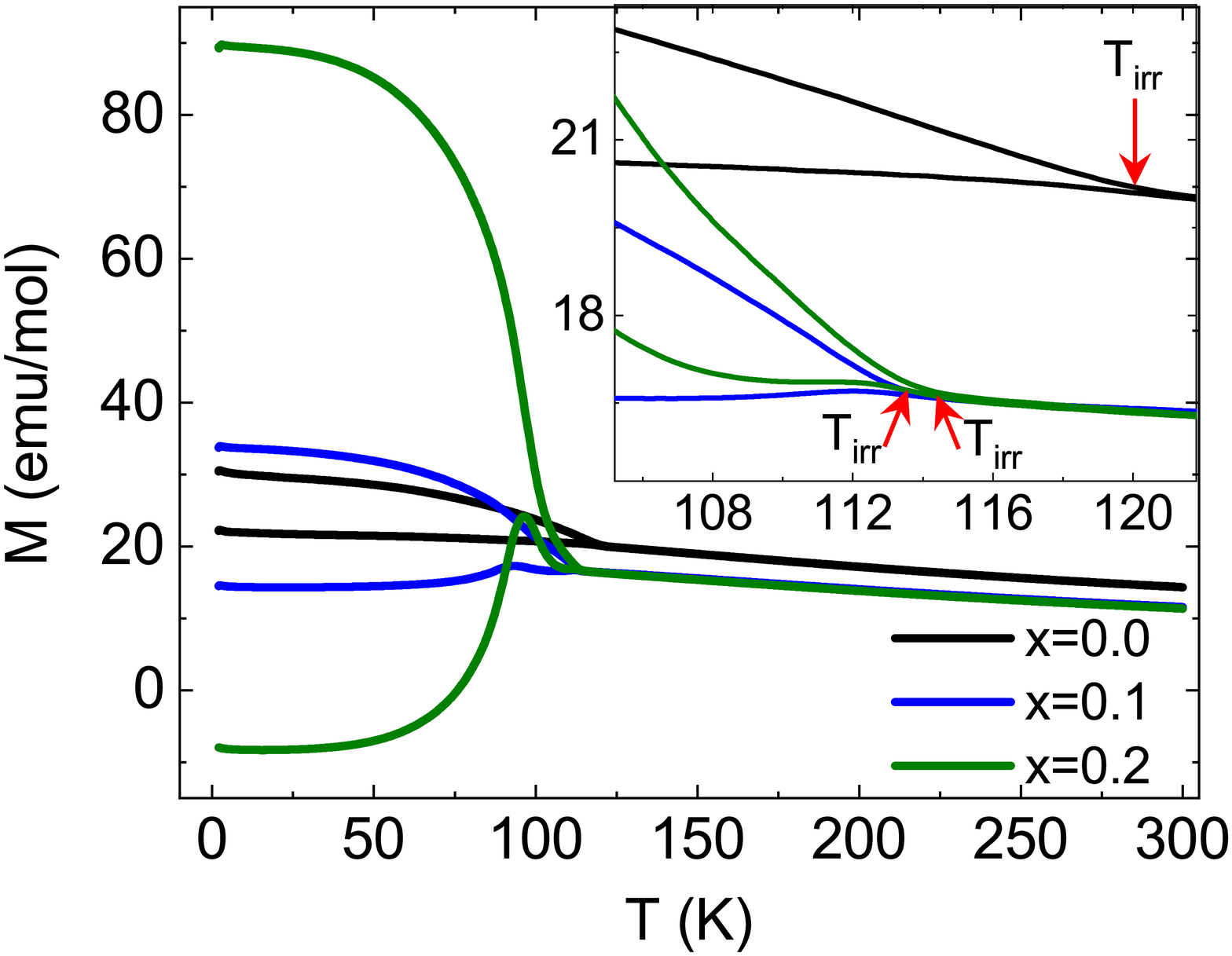}
	\caption{Temperature dependent magnetization measured at 1 kOe in ZFC FC protocol for Eu$_{2}$(Ir$_{1-x}$Cu$_{x}$)$_{2}$O$_{7}$, where $x$=0, 0.1, 0.2. Inset: Enlarged view of the irreversibility temperature.}                    
	\label{MT}
\end{figure} 

Fig. \ref{MT} represents temperature dependent magnetization (M) for all the samples under zero field cooled (ZFC) - field cooled (FC) protocol with applied field 1 kOe. Parent sample shows a magnetic irreversibility around T$_{irr}$ = 120 K, below which there is a bifurcation between ZFC and FC curve. Such type of magnetic behavior was reported in the literature.\cite{Ishikawa} From  resonant x-ray diffraction (RXD) and muon spin rotation and relaxation ($\mu$SR) studies, it was revealed that in this compound below T$_{irr}$, Ir moments order in an antiferromagnetic all-in/all-out spin arrangement.\cite{Zhao, Sagayama} We observe that T$_{irr}$ for $x$=0, 0.1 and 0.2 compounds are near about 120 K, 113 K, 115 K respectively (see inset in fig. \ref{MT}). Therefore, all the compounds undergo metal insulator transition almost at their respective irreversibility temperature. There exhibits a prominent cusp in the ZFC magnetization around T$_{irr}$ for all the compounds.  We observe that the difference between M$_{ZFC}$ and M$_{FC}$ is systematically increases with Cu doping. The ZFC magnetization below T$_{irr}$ gradually shifts to lower value with doping (Note: the negative magnetization observed in the ZFC for $x$=0.2 is due to the small negative field trap in the instrument while cooling in zero field condition and is attributed to an instrumental artifact). Since each Cu ion creates two Ir$^{5+}$ non-magnetic ions, as a result, the ZFC magnetization of the system decreases with doping. On the other hand, the FC magnetization is found to increase with Cu doping. This further indicates increase in field induced magnetization with increase in doping concentrations. Bifurcation is observed due to formation of field induced ferro correlation within the droplets containing Cu with the applied magnetic filed (i.e. in those regions where AIAO deviates due to Cu$^{2+}$ doping). These droplets are dispersed in
an AIAO AFM background matrix and the flat plateau observed at very low temperature indicates that these ferro droplets are randomly arrested and frozen in the AIAO AFM background matrix. With increase in temperature, these super-moments tend to unblock giving rise to the observed bifurcation below T$_{irr}$.

\begin{figure} 
	\centering
	\includegraphics[width= 8 cm]{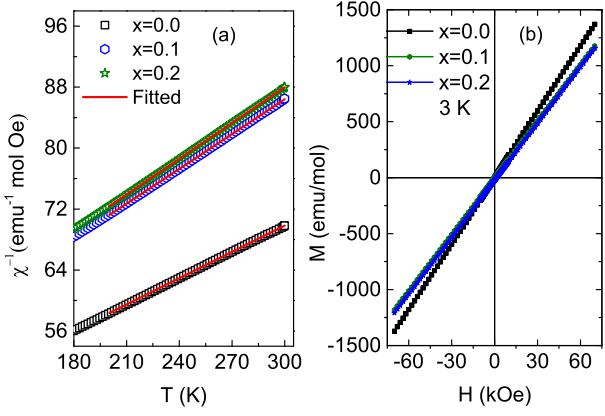}
	\caption{(a) Temperature dependent inverse susceptibility with Curie Weiss fit. (b) Isothermal magnetization at 3 K for  Eu$_{2}$(Ir$_{1-x}$Cu$_{x}$)$_{2}$O$_{7}$, where $x$=0, 0.1, 0.2.}
	\label{Invchi-T_M-H}
\end{figure}
\begin{table}
	
	\centering \caption{$\mu_{eff}$, $\theta_{p}$ and $f$ for Eu$_{2}$(Ir$_{1-x}$Cu$_{x}$)$_{2}$O$_{7}$ system.}

	\begin{tabular}{cccc}
		\hline
		\hline Sample&\hspace{0.5cm}$\mu_{eff}$ ($\mu_{B}$/f.u.)&\hspace{0.5cm}$\theta_{p}$ (K)& \hspace{0.5cm}$f$\\
		\hline $x$=0 & 8.32 &\hspace{0.4cm}-306.2$\pm$0.2&\hspace{0.8cm}2.551$\pm$0.002\\
		\hline $x$=0.1 &7.24 &\hspace{0.4cm}-267.4$\pm$0.1&\hspace{0.8cm}2.3663$\pm$0.0009\\
		\hline $x$=0.2 &7.14 &\hspace{0.4cm}-262.8$\pm$0.1&\hspace{0.8cm}2.2852$\pm$0.0009\\
		\hline
		\hline
	\end{tabular}
\label{chifittab}
\end{table}

Fig. \ref{Invchi-T_M-H}(a) shows temperature dependent inverse susceptibility for $x$=0, 0.1, 0.2, fitted in the temperature region 200-300 K with the Curie Weiss law  $\chi = \frac{C}{T-\theta{p}}$, $\theta{p}$ is  Curie Weiss  temperature, Curie constant  
$C= \frac{N_A\mu_{eff}^{2}}{3K_{B}}$, and  $\mu_{eff}$ is the effective paramagnetic moment. The fitted parameters $\mu_{eff}$, $\theta_{p}$ and frustration parameter ($f = |\theta_p|/T_{irr}$, indicates the level of frustration in the system) are listed in table \ref{chifittab}. All the samples show negative $\theta_{p}$, which indicates antiferromagnetic interaction, whose strength decreases with Cu doping. We also observe that frustration parameter $f$ decreases with Cu doping, which signifies the decrease in frustration with Cu concentration. From XPS measurement, we observed that $x$ amount of Cu$^{2+}$  creates 2$x$ amount of Ir$^{5+}$, hence the effective moment can be expressed as:
\begin{equation} 
\mu_{eff}=\sqrt{(1-3x)(Ir^{4+}_{\mu_{eff}})^2+2x(Ir^{5+}_{\mu_{eff}})^2+x(Cu^{2+}_{\mu_{eff}})^2}.
\end{equation}

 As Ir$^{5+}$ is non-magnetic, the effective moment decreases with Cu concentration.
  
 Isothermal magnetization curve (M-H) for all the samples measured at 3 K up to field 70 kOe are shown in fig. \ref{Invchi-T_M-H}(b). All the compounds show linear behavior without any hysteresis. Magnetic moment at highest applied magnetic field (70 kOe) decreases with Cu doping. The reduction of moment with Cu doping is due to the creation of non-magnetic Ir$^{5+}$ ions.
\begin{figure} 
	\centering
	\includegraphics[width= 8 cm]{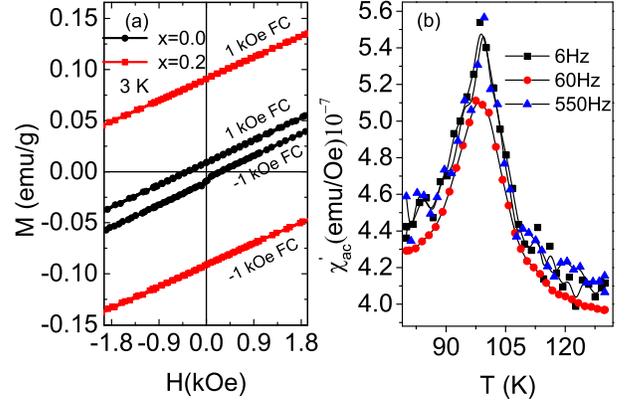}
	
	\caption{(a) Isothermal magnetization at 3 K after the sample was cooled in presence of magnetic field 1 kOe and -1 kOe for Eu$_{2}$(Ir$_{1-x}$Cu$_{x}$)$_{2}$O$_{7}$, where $x$=0, 0.2. (b) Temperature dependent AC susceptibility for $x$=0.2 compound with different frequencies.}
	\label{FC_MH}
\end{figure}

Magnetic field cooled M-H were measured at 3 K after the samples were cooled down from room temperature in presence of a magnetic field 1 kOe and -1 kOe. From fig. \ref{FC_MH}(a), we observe a shift of the M-H curve along the magnetization axis for $x$=0 and 0.2 compounds. The direction of the shift depends on the polarity of the field. The magnitude of the shift increases with Cu doping though there is no hysteresis loop for both the compounds. Such type of cooling field induced shift of the M-H curve has been previously observed in the pyrochlore iridates.\cite{Yang} Since each corner shared tetrahedra deviates from its all-in-all-out spin structure and it picks up a moment in both the samples ($x$=0 and 0.2) due to creation of non-magnetic Ir$^{5+}$ state, leading to a small ferro contribution. This ferro contribution interacts with background AIAO AFM spin matrix, giving an additional exchange field due to pinning with the ferro spin and produce exchange bias like behavior. The size and amount of the ferro island/droplets increases with increase in Cu concentration. This leads to larger amount of ferro island pinning to the AIAO AFM background and hence, increases the shift (exchange bias) of the M-H curve with increase in Cu doping.

AC magnetization measurement is an efficient tool to detect the existence of non-equilibrium phase in the compound. We have carried out temperature dependent AC susceptibility measurement for $x$=0.2 compound at different frequencies as shown in fig. \ref{FC_MH}(b). A peak at T= 95 K was observed and does not shift with frequency. This confirms that there is no glassy-like dynamics below the T$_{irr}$ in the doped compound and further reflects the AFM phase transition.
\begin{figure} 
	\centering
	\includegraphics[width= 9 cm]{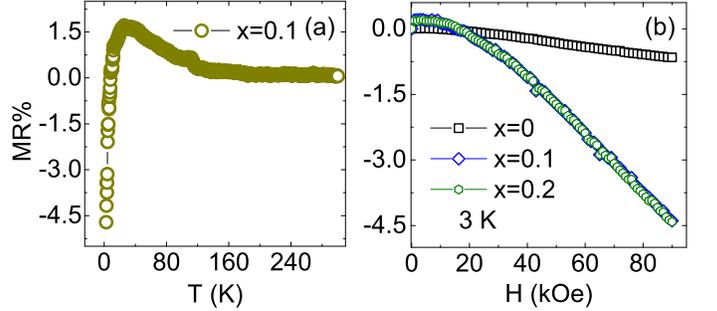}
	\caption{(a) Temperature dependent magnetoresistance for $x$=0.1 compound. (b) Magnetic filed dependent magnetoresistance at 3 K for  Eu$_{2}$(Ir$_{1-x}$Cu$_{x}$)$_{2}$O$_{7}$, where $x$=0, 0.1 and 0.2.}
	\label{MR}
\end{figure}
  
  Magneoresistance (MR), is defined by ($\rho_H-\rho_0)/\rho_0$, where $\rho_H$ and $\rho_0$ are the resistivity with field and without field. Fig. \ref{MR}(a) shows temperature dependent MR for $x$=0.1 compound. We observe that below T$_{MI}$ MR is negative. MR as a function of magnetic field measured at 3 K (Fig. \ref{MR}b) shows negative value for all the samples. Magnitude of the negative MR increases with Cu doping due to increase of field induced ferro contribution regions where the spin-flip scattering is lower than that of the AIAO AFM structure.
  
   \begin{figure} 
   	\centering
   	\includegraphics[width= 9 cm]{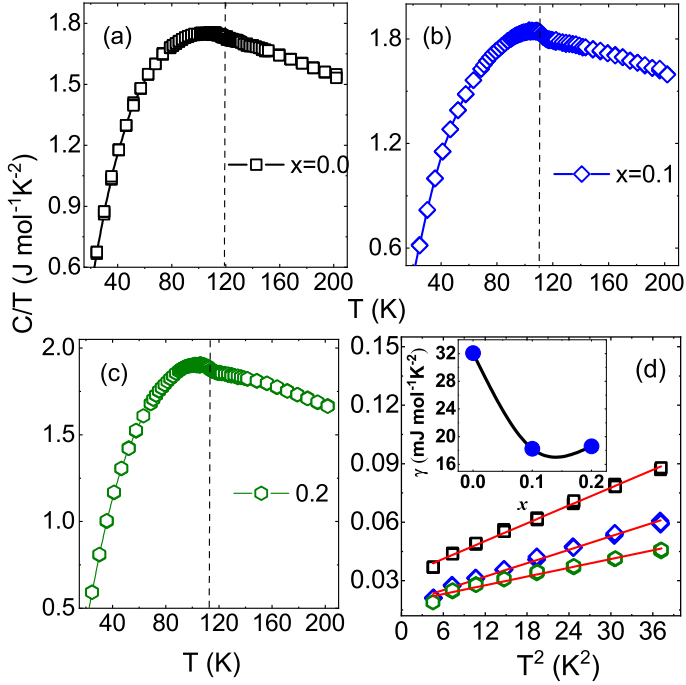}
   	\caption{(a-c) Plot of temperature dependent specific heat (C/T) in the temperature range 20-200 K for the samples Eu$_{2}$(Ir$_{1-x}$Cu$_{x}$)$_{2}$O$_{7}$. (d) Representation of C/T Vs. T$^2$ plot for all the samples at low temperature. Solid red lines are fitted line. Inset: Variation of the electronic contribution ($\gamma$) with Cu concentration.}             
   	\label{HC}
   	\end{figure}
   
   Fig. \ref{HC}(a-c) shows specific heat data as a function of temperature for all the samples in the temperature range 20-200 K. Vertical dotted line in C/T Vs. T data for all the samples marks the expected anomaly. We do not observe any anomaly for $x$=0 but clearly observe for $x$= 0.1 and 0.2 at 110 K and 112 K respectively. This anomaly temperature is close to the magnetic irreversibility temperature (T$_{irr}$). To determine the change of electronic contribution $\gamma$, we have fitted C/T Vs. T$^2$ data with the formula 
   
     \begin{equation}
   C = \gamma T + \beta T^3
   \end{equation} 

   in the temperature region 2.1-6 K for all the compounds, as shown in fig. \ref{HC}(d).  We obtain the $\gamma$ value for the parent compound to be 32 mJ mol$^{-1}$ K$^{-2}$ which is consistent with the value reported by Ishikawa et al.\cite{Ishikawa} From the inset of fig. \ref{HC}(d) we observe that $\gamma$ abruptly decreases with Cu doping.  
   
   \vskip 0.5 cm
    {\large\bf{Discussion}} 
   \vskip 0.2cm
   In the parent Eu$_{2}$Ir$_{2}$O$_{7}$ compound, the transport property is driven by the $5d$ transition metal Ir. The transition metal is subjected to a trigonal distortion of the octahedral field. The ionic radii\cite{Barsoum} of Cu$^{2+}$, Ir$^{4+}$ and Ir$^{5+}$ are 0.87 \AA, 0.765 \AA, and 0.71 \AA$\space$ respectively with coordination number six. As the ionic radius of Cu$^{2+}$ is greater than Ir$^{4+}$ each Cu at Ir site tends to increase the volume of the unit cell. On the other hand, each Cu will convert two nearby Ir$^{4+}$ sites into Ir$^{5+}$, thereby contracting the unit cell. The overall effect is a reduction of lattice spacing with increasing Cu concentrations.  
   When each Cu gives rise to two Ir$^{5+}$ ions, this is like hole doping in a pyrochlore spin 1/2 antiferromagnet. This could lead to the reduction in T$_{MI}$, which was observed in the systems where Sr replaces Eu.\cite{Banerjee} However, we did not see any significant change in T$_{MI}$, while Cu concentration increases from 10\% to 20\%. We believe that energy level available at Cu site for an incoming electron is higher than the energy level of the electron at the Ir site. Hence Cu$^{2+}$ sites are not available for transport of d-electrons on the Ir sublattice. These sites are effectively blocked for transport (though they retain their spin) and resistivity increases with Cu doping. The situation can be regarded as a percolation problem where sites are being removed randomly. T$_{MI}$ is insensitive to Cu doping for concentration more than 10 \% since it is beyond the 3-d percolation limit.  
   
    At low temperature region (below 50 K) the resistivity data follows power law like variation. Below T$_{MI}$ the rise in Seebeck coefficient indicates the increase of insulating nature of the samples with Cu doping. A significant negative Hall voltage appeared at low temperature can be associated to a spin fluctuation influenced by the magnetic field. In this temperature region the dominant charge carrier for all the compounds is electron-like, where carrier concentration decreases with Cu doping.  
  
   Both Ir and Cu have pseudo spin 1/2 single ion anisotropy but their orbitals are different. The ground state spin arrangement of Ir is all-in-all-out (AIAO). There is an unavoidable non-stoichiometry in the parent compound\cite{Ishikawa,Zhu}, creates non-magnetic Ir$^{5+}$ state, which deviates the spin structure from AIAO. Each tetrahedra picking a moment in an applied field forms a field induced ferro droplets, which pinned within the background AIAO AFM spin. This further gives rise to exchange bias like behavior and increases with Cu doping due to increase of Ir$^{5+}$ state. These ferro droplets are randomly frozen in the AIAO AFM background matrix giving rise to the bifurcation in ZFC-FC magnetization depending on the amount and size of the droplets and it increases with Cu concentration. 
  
    For the parent compound MR is negative at low temperature with small magnitude, which varies quadratically at low field and linearly at high field. We proposed such type of MR behavior is due to the chiral spin liquid.\cite{Mondal,Fujita,Machida} Increase in Cu concentration reduces the AIAO AFM region which in turn reduces the region of  spin liquid phase as indicated in the specific heat data. The doping of Cu increases the field induced ferro correlated region and consequently the MR increases due to reduction of spin-flip scattering from the ferro droplet region.  
   
   Now we would like to mention another important observation, the existence of linear specific heat for all the samples at low temperature. Generally, the linear specific heat at the insulating state arises from the weak localization due to finite density of state at the Fermi level. But in our samples, resistivity data rules out the weak localization nature of the resistivity at the insulating state. On the other hand, the antiferromagnetic frustrated  lattice often gives rise to spin liquid ground state, which supports fractional excitation, called spinon.\cite{Balents} Spin liquid is a kind of spin super fluid (pairing between spinon with opposite spin), where spin excitations occur with or without a gap.\cite{Baskaran} Theoretical study of an antiferromagnetic kagome lattice indicates that a linear specific heat at the insulating state can be possible if the spinons have finite area Fermi surface.\cite{Ran} Since Cu doping creates regions (droplets) of  deviated AIAO structure and thus, with increase in Cu doping the prerequisite condition for spin liquid formation region decreases, which is observed in the specific heat data. We also notice the onset of a negative Hall voltage in the temperature region where the spin liquid state exists. This may be due the spin fluctuation in the spin liquid state, and needs to be addressed furthermore by theory.

  \vskip 0.5cm
   {\large\bf{Conclusion}}
   \vskip 0.2cm
  From our analysis of  magnetic, electrical and thermal behavior of Cu doped pyrochlore iridate Eu$_{2}$(Ir$_{1-x}$Cu$_{x}$)$_{2}$O$_{7}$, we conclude that each Cu creates two Ir$^{5+}$ which deviates each corner shared tetrahedron from its AIAO spin structure and picks up a magnetic moment. It does not support the spin liquid state in the vicinity of Cu within a background of antiferromagnetically coupled AIAO spin structure. Observation of linear specific heat at low temperature indicates quantum spin fluctuation of spinons. With increasing Cu concentration, the coefficient of the linear specific heat decreases indicating reduction of spinon density which is attributed to the reduction of the AIAO AFM region. Increase in resistivity below T$_{MI}$ is due to onset of AFM phase transition, and power law dependent increase in resistivity with lowering temperature indicates that it is neither an activated nor Mott variable range hopping as expected in 3-dimensional disordered system. At low temperatures, increase of resistance and sudden development of negative Hall voltage can be associated with the onset of spin liquid formation. The doping of Cu increases the field induced ferro correlated region and as a result the MR increases due to reduction of spin-flip scattering.  The bifurcation observed in the ZFC-FC magnetization which increases with Cu doping is due to formation of randomly oriented blocked supermoments embedded in AIAO AFM phase. Increase in exchange bias with Cu doping is due to the increase in amount/size of pinning of the ferro spin within AFM AIAO matrix of the system.       
 \vskip 0.5cm
   {\large\bf{Acknowledgements}}
   \vskip 0.2cm
   S.M. would like to thank Prof. Manabendra Mukherjee, Prof. Satyajit Hazra, Mr. Goutam Sarkar, SPMS Division, SINP for XPS measurement and Dr. Mrinmay K Mukhopadhyay, Dr. Ramkrishna Dev Das, SPMS Division, SINP for XRD measurement. S.M. acknowledges UGC-DAE CSR, Kolkata for Hall measurement. Authors acknowledge CIF, IISER Bhopal, for PPMS and SQUID-VSM facilities. This work is partially supported by SERB, DST, GOI under TARE project (File No.TAR/2018/000546).

\end{document}